\documentclass[sigconf]{acmart}
\usepackage{amsmath,amsfonts}
\usepackage{algorithmic}
\usepackage{graphicx}
\usepackage{textcomp}
\usepackage{xcolor}
\usepackage{subcaption}
\usepackage{enumitem}
\usepackage{multirow}
\usepackage{hyperref}
\usepackage{balance}
\usepackage{booktabs}
\AtBeginDocument{%
  \providecommand\BibTeX{{%
    \normalfont B\kern-0.5em{\scshape i\kern-0.25em b}\kern-0.8em\TeX}}}

\setcopyright{acmcopyright}
\copyrightyear{2023}
\acmYear{2023}
\acmDOI{XXXXXXX.XXXXXXX}

\acmConference[EASE 2023]{The 24th International Conference on Evaluation and Assessment in Software Engineering}{14–16 June, 2023}{Oulu, Finland}
\begin{document}

\title{DQSOps: Data Quality Scoring Operations Framework for Data-Driven Applications}


\author{Firas Bayram}
\orcid{0000-0003-0683-2783}
\email{firas.bayram@kau.se}
\affiliation{%
  \institution{Dept of Mathematics and Computer Science, Karlstad University}
  \city{Karlstad}
  \country{Sweden}
  \postcode{651 88}
}

\author{Bestoun S. Ahmed}
\orcid{0000-0001-9051-7609}
\email{bestoun@kau.se}
\affiliation{%
  \institution{Dept of Mathematics and Computer Science, Karlstad University}
  \streetaddress{Universitetsgatan 2}
  \city{Karlstad}
  \country{Sweden}
  \postcode{651 88}
  \institution{\\Department of Computer Science, Czech Technical University}
  \city{Prague}
  \country{Czech Republic}
}
% \affiliation{%
%   \institution{Department of Computer Science,\\Czech Technical University}
%   \streetaddress{Universitetsgatan 2}
%   \city{Karlstad}
%   \state{V{\"a}rmlands l{\"a}n}
%   \country{Sweden}
%   \postcode{651 88}
% }

\author{Erik Hallin}
\email{erik.hallin@uddeholm.com}
\author{Anton Engman}
\email{anton.engman@uddeholm.com}
\affiliation{%
  \institution{Uddeholms AB}
  \streetaddress{Uvedsv{\"a}gen}
  \city{Hagfors}
  \state{V{\"a}rmlands l{\"a}n}
  \country{Sweden}
  \postcode{683 33}
}

\renewcommand{\shortauthors}{Bayram F, et al.}

\begin{abstract}
Data quality assessment has become a prominent component in the successful execution of complex data-driven artificial intelligence (AI) software systems. In practice, real-world applications generate huge volumes of data at speeds. These data streams require analysis and preprocessing before being permanently stored or used in a learning task. Therefore, significant attention has been paid to the systematic management and construction of high-quality datasets. Nevertheless, managing voluminous and high-velocity data streams is usually performed manually (i.e. offline), making it an impractical strategy in production environments. To address this challenge, DataOps has emerged to achieve life-cycle automation of data processes using DevOps principles. However, determining the data quality based on a fitness scale constitutes a complex task within the framework of DataOps. This paper presents a novel Data Quality Scoring Operations (DQSOps) framework that yields a quality score for production data in DataOps workflows. The framework incorporates two scoring approaches, an ML prediction-based approach that predicts the data quality score and a standard-based approach that periodically produces the ground-truth scores based on assessing several data quality dimensions. We deploy the DQSOps framework in a real-world industrial use case. The results show that DQSOps achieves significant computational speedup rates compared to the conventional approach of data quality scoring while maintaining high prediction performance.
\end{abstract}




\maketitle
\section{Introduction}
In the era of big data explosion, data has become the most critical asset of artificial intelligence (AI) software projects \cite{meng2013big}. The primary motivation is that the data accompany machine learning (ML) software throughout their life cycle. Notably, huge volumes of data are being collected for industries and businesses at an ever-escalating rate. The volume and velocity pose a formidable challenge for ML software systems in production, which are required to make (near) real-time decisions \cite{teh2020sensor}. In real-life scenarios, the challenge is two-fold: assessing the quality of the data flowing in the system and processing it within defined time frames.

%%% Bestoun 10 January 2023
Data quality is not a new topic; its roots are traced back to the literature on database management systems \cite{wang1995framework}. However, the advancement of data-driven systems has recently shifted the topic towards AI research \cite{batini2015data}. It is widely recognized that the performance of ML projects is mainly dependent on the quality of the underlying data \cite{jain2020overview}. Data quality is a multifaceted concept that defines several quality dimensions based on the relative nature of the problem \cite{moges2013multidimensional}. Typically, data quality dimensions are identified according to the context and domain, which implies that data quality dimensions may change with application. Each dimension of the data quality measures the condition of the data concerning a specific aspect which is usually summarized in a quality index or metric that is used as an indicator of data validity \cite{vaziri2019measuring}. 

Data quality scoring holds significant relevance in data quality assessment and is defined as the methodology for obtaining a score metric for each dimension of the data quality of a given set of observations in a dataset \cite{taleb2021big}. In the data quality scoring literature, researchers investigate the procedures to measure and quantify the different dimensions of data quality within the target application. The scores are interpreted as the fitness scale corresponding to certain data quality dimensions for each data record or chunk. Data quality scores are used in data storage management or leveraged in predictive ML software by introducing acceptability thresholds \cite{loshin2010practitioner}. Acceptability thresholds filter out data instances that do not meet quality standards. As a result, the data quality scoring methodology enables the characterization of data records according to their quality through quantitative analysis.

%%% Bestoun 10 January 2023

The conventional procedure for data quality scoring is executed through a code that manually inspects and validates the condition of the data with respect to the defined data quality dimensions \cite{sunderland2019utility}. However, this traditional procedure is resource intensive, especially if the number of data quality dimensions is high, and generally requires human intervention \cite{muller2019data}. Furthermore, in deployment environments, this manual procedure is not practical for ML software that should perform tasks automatically \cite{ledell2020h2o}. Recently, the DevOps principles, continuous integration and continuous delivery (CI/CD), have been adopted and introduced to build productionized ML software, known as Machine Learning Operations (MLOps) \cite{john2021towards}. MLOps systems allow the deployment of ML software in a formalized way throughout the life-cycle of AI applications \cite{bosch2021engineering}. An indispensable element of MLOps is DataOps, which is the approach that manages and processes the production data systematically and continuously. DataOps pipelines are designed to prepare accurate and reliable data that the ML model will use. However, scoring the streaming data windows is challenging and yet to be explored within the context of DataOps for production systems that are characterized by high sampling rates.

This paper contributes to the current advancement in the DataOps field by presenting a novel Data Quality Scoring Operations (DQSOps) framework that can be streamlined in DataOps workflows. DQSOps can be viewed as a general data scoring methodology to automate scoring the quality of streaming data. The generalizability of DQSOps is enabled by incorporating configurable components that can be easily adjusted and extended based on the problem. In practical implementations, DQSOps accelerates scoring the quality of acquired production data irrespective of the number of defined quality dimensions. The speedup is achieved by employing an innovative ML predictor that determines the quality score of the processed data window. Using an ML predictor rather than the standard data scoring methodology substantially reduces the overall processing time. However, a test oracle is implemented to continuously evaluate the ML accuracy throughout the system evolution to sustain high performance for the ML predictor. The test oracle uses ground-truth data quality scores that are periodically produced by a standard-based approach. Additionally, we introduce a new data mutation component, inspired by the mutation testing principles, to simulate data quality issues in the data window, and thus facilitate the initialization phase of our framework.

The rest of the paper is structured as follows: Section \ref{sec:background} reviews the related work in the field and provides an overview of the data quality scoring task. Section \ref{sec:methodology} presents the methodology used to develop our data quality scoring framework. The experimental results of our use cases are reported in Section \ref{sec:results}. The threats to generalize our proposed framework are discussed in Section \ref{sec:threats}. Finally, Section \ref{sec:conclusion} concludes the paper with a summary of the remarks and future perspective.

%%% Bestoun 10 January 2023

\section{Background and Previous Work}
\label{sec:background}
This section provides a general overview of data quality dimensions and related definitions. Then we review studies devoted to assessing data quality in the literature.

\subsection{Data Quality Dimensions}
\label{sec:dqd}
Data quality dimensions are defined to assess the quality of the data. Each dimension captures a specific characteristic of the evaluated data within a particular task. Knight \cite{knight2011combined} has presented a combined conceptual framework for the quality of information systems. ML software concerns the relation between the system’s data and task attributes in information systems. Therefore, contextual quality is relevant to build an ML solution, as it is identified in the interactive space between the data of a system and the attributes of the task \cite{dedeke2000conceptual}.

The contextual quality comprises five quality dimensions: \textit{value-added}, \textit{relevancy}, \textit{timeliness}, \textit{appropriate data} and \textit{completeness} \cite{knight2011combined}. However, since these quality dimensions are context-based, they must be adapted to the task and the system's characteristics. Wand and Wang \cite{wand1996anchoring} described the dimensions of data quality as subjective and should be defined within the application of the system that generates the data. In a more recent study, \cite{taleb2021big}, these dimensions have been revised for continuous quality management. The revised data quality dimensions are: \textit{accuracy}, \textit{completeness}, \textit{consistency}, \textit{Timeliness}. However, for evolving ML software systems, distributional changes are likely to occur during system evolution \cite{lu2018learning}. Therefore, we will extend these quality measures and include \textit{skewness} \cite{chug2021statistical} to monitor the distribution of incoming data. Descriptions of the data quality dimensions are provided with the following definitions:

\begin{enumerate}[label={\arabic*.}] 
    \item \textbf{Accuracy:} Measures whether the observed data value represents the actual value.
    \item \textbf{Completeness:} Checks whether the observed data include missing values.
    \item \textbf{Consistency:} Verifies whether the observed values meet the integrity constraints of the domain.
    \item \textbf{Timeliness:} Describes whether the data are up-to-date for the corresponding task. 
    % currency
    \item \textbf{Skewness:} Computes the distribution deviation of the observed data from a reference distribution. 
\end{enumerate}

% \label{sec:background}
% In this section, we review the studies devoted to assessing data quality in the literature. We then give a general overview of the data quality dimensions and related definitions.

\subsection{Related Work}
Data quality assessment has been the subject of many early works in the literature \cite{olson2003data}. Chug \textit{et al.} \cite{chug2021statistical} have proposed a method that quantifies the quality of a given dataset. The method uses nine data quality dimensions to estimate the quality of the dataset. As a result, the method provides a score, report, and label for the data quality of the dataset. Similarly, for big data systems, Taleb \textit{et al.} \cite{taleb2021big} proposed the Big Data Quality Management Framework (BDQMF) to address data quality issues in big data systems based on several data quality dimensions. The framework included several components to manage, validate, and monitor the data quality. Furthermore, data quality issues were investigated both at the cell instance and the schema levels of the dataset. BDQMF framework also quantifies scores of data quality aspects. However, the authors did not experimentally evaluate the framework. For business processes, Data Quality Validation Methodology (DQVM) was proposed to evaluate the effects of data quality on the process outcome \cite{cappiello2018validating}. The methodology allows domain experts to assign the corresponding weights for each data quality dimension based on their relative importance to the business process. The deviation of quality scores is then observed between fault-free and fault-injected datasets, and the impact on the process is eventually evaluated.

Other approaches have investigated quantifying an individual data quality dimension. Heinrich \textit{et al.} \cite{heinrich2018assessing} proposed a quality metric based on probability theory to assess \textit{semantic consistency}. The metric indicates the degree to which assessed data is contradiction-free. Similarly, \textit{minimality}, or \textit{uniqueness}, was investigated to measure redundancies in data at the data- and schema-level \cite{ehrlinger2018novel}. The method uses hierarchical clustering and similarity calculation techniques to calculate the metric. For Internet of Things (IoT) systems, Byabazaire \textit{et al.} \cite{byabazaire2022end} presented a real-time data quality assessment framework. To measure data quality, the framework uses \textit{trust} as a metric for quantification. The correlation between trust score and root mean square error (RMSE) and mean absolute error (MAE) is calculated using Pearson's correlation coefficient to validate if trust is a good indicator of data quality. With the limited approaches to address data quality scoring in the context of production systems, we design a framework that can be utilized efficiently in real-world problem scenarios. The proposed framework can handle streaming data with high sampling rates, as opposed to the existing methods that are designed for offline static datasets. Additionally, our proposed framework can effectively accommodate a high number of quality dimensions.

%%% Bestoun 11 January 2023

\section{Data Quality Scoring Operations (DQSOps) Framework}
This section introduces the methodology to deliver our proposed DQSOps framework. DQSOps can efficiently evaluate the quality of production data streams based on quantitative analysis. The output of DQSOps is a score metric that serves as an indicator of the overall validity of the collected data window, and is calculated based on assessing several data quality dimensions. The scores can be used to rank the system data according to their quality. Thus they can be selected for the various business analysis tasks that may require different levels of data quality. In practice, DQSOps significantly reduces the time to score the data window by employing an ML predictor that replaces the conventional data scoring method. The run time is irrespective of the number of quality dimensions, making it convenient for real-world applications characterized by high sampling rates. In addition, the performance of the ML predictor is periodically evaluated by executing test oracles. 

\begin{figure*}
    \centering
    \includegraphics[scale=0.9]{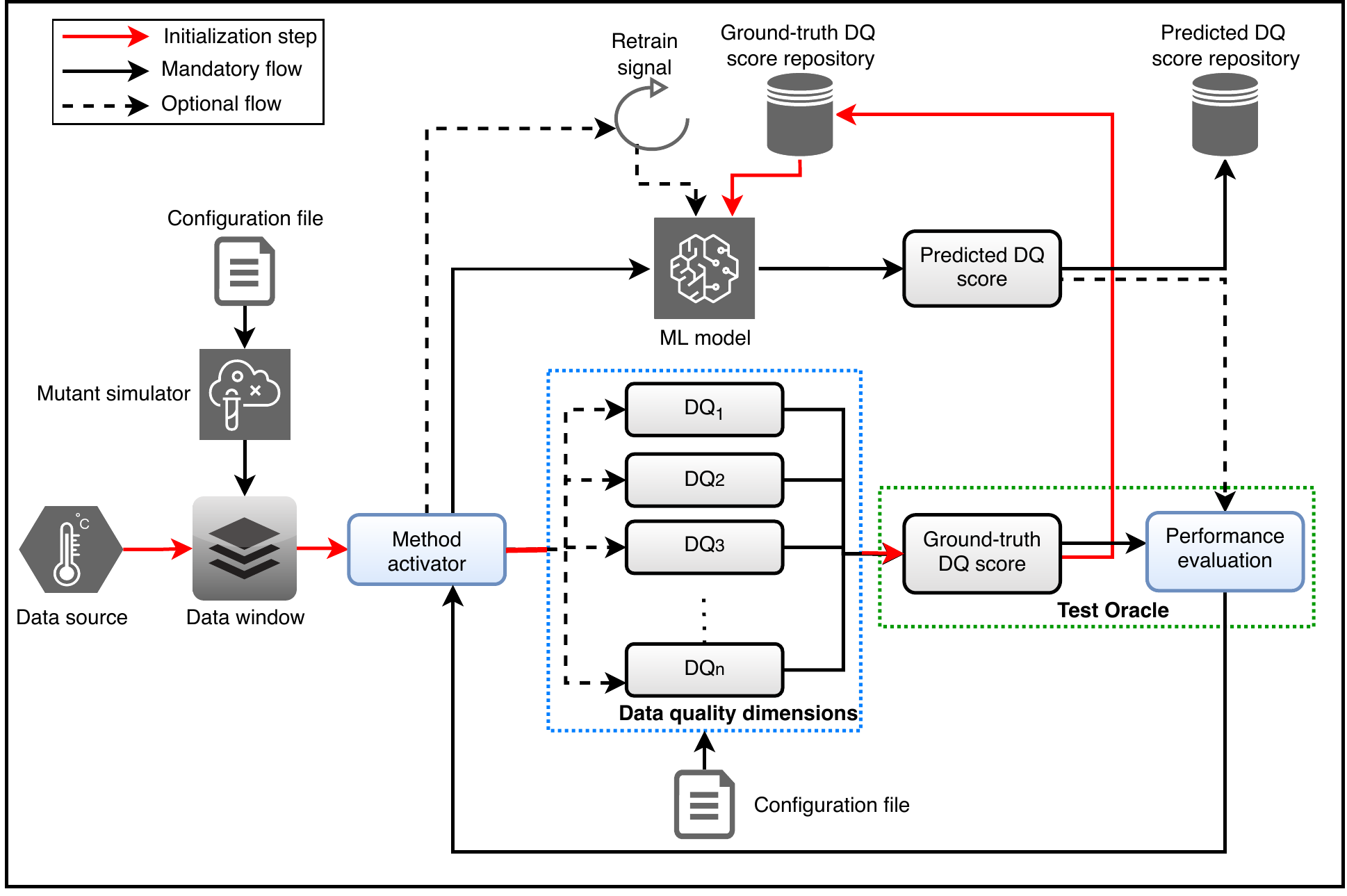}
    \caption{DQSOps framework, the red path represents the initialization phase.}
    \label{fig:DSOps}
\end{figure*}

The overall pipeline of the presented framework is shown in Figure~\ref{fig:DSOps}. Upon receiving the data streams from data-generating sources, the stream is segmented into windows of data samples according to a pre-defined data window size. Inspired by the mutation testing principles \cite{jia2010analysis}, we introduce a \textit{data mutant simulator} component to obtain full control over the experimental conditions. The component is applied to simulate \textit{nonequivalent data mutants} within the data window according to pre-configured parameters that are loaded from a configuration file that stores the pre-specified mutation percentage of each data quality dimension. Nonequivalent mutants are used to induce meaningful changes in the problem \cite{schuler2013covering}. Typical examples of data mutants are missing or inconsistent data values, or anomalies. Consequently, we utilize the data mutation component to simulate erroneous data that would affect the data quality according to the specified dimensions. Furthermore, the component is especially useful during the initialization phase, as will be discussed further in Section \ref{sec:init}.
% This component gives complete control over the and 
% fault rates since we can perturb the original data by a certain ratio with respect to specific data quality dimension(s). The ratio settings are loaded from a configuration file that stores the perturbation rate for each data quality dimension. 

\label{sec:methodology}

After preparing the data window, the method activator component is invoked to activate the appropriate workflow path for data scoring based on particular decisions. The criteria for selecting the approach will be further detailed in Section \ref{sec:activ}. The candidate quality scoring approaches are: the standard scoring method, a regression ML model, or a retrain signal. The standard scoring method calculates the ground-truth data quality scores of the defined dimensions and stores it in a repository. Whereas the ML model predicts the data quality score of each data window. If the retrain signal is activated, the ML model will be retrained using the ground-truth scores stored in the repository. During the retraining process, both the current prediction- and standard-based approaches will be used to score the data quality until the new model is ready to replace the current one. Furthermore, the performance of the regression model is continuously monitored using a test oracle that evaluates the predicted data quality scores. The next subsections will delve deeper into the mechanisms of the main components of the DQSOps framework.

% In the next subsections, we first elaborate on the strategy for quantifying a score of the different data quality dimensions. We then show how we aggregated the calculated scores into a consolidated score that represents the data quality of each data record.

\subsection{Scoring the Data Quality Dimensions}
\label{sec:scoring}
The standard-based approach is a core element of DQSOps and used to quantify the ground-truth scores of the data quality dimensions that are used to train the ML model. Each data quality dimension will be reflected in one quality score. In particular, this score quantifies the level of fitness of the collected data window with respect to the specific data quality dimension. The relevant data quality dimensions are: accuracy, completeness, consistency, timeliness, and skewness, as discussed in Section \ref{sec:dqd}. As shown in Figure \ref{fig:DSOps}, meta-information is loaded that helps calculate the scores of the data quality dimensions from a configuration file. For instance, meta-information may include the values that characterize the integrity constraints of the specific problem, such as the maximum and minimum value limits. The configuration file could also include the path of auxiliary files that supports the calculation of some scores such as the anomaly detector or data distribution. After calculating the data quality scores, all score values are standardized between 0 and 1 using min-max normalization to obtain uniform scales across the data quality dimensions, as discussed in the next section. The data quality scores are retrieved as follows:

\subsubsection{Accuracy Score}
Accuracy score is calculated by finding the proportion of the anomalous datum in the processed data window \cite{taleb2021big}: $Accuracy=\frac{NAV}{N}$, where $NAV$ is the total number of anomalous values detected in the data window and $N$ is the size of the data window. Anomalies may appear in the system due to several factors, such as malicious activities, hardware failures, inaccuracies in data collection, or adversarial attacks \cite{chalapathy2019deep}. From a data quality perspective, anomalous data records represent abnormal values of unhealthy data instances and therefore are considered an indicator of low-quality data \cite{rettig2019online}. The anomaly detector can be obtained in the initialization step; see Section \ref{sec:init}.

\subsubsection{Completeness Score}
The fraction of missing values that are observed in the data window and can be found as follows \cite{heinrich2018requirements}: $Completeness=\frac{NNV}{N}$, where $NNV$ is the number of missing values such as NA (Not Available) or NULL observations and $N$ is the size of the data window. Missing values are a popular problem of data quality that could be an indicator of disconnection or damage to the data source \cite{peng2005review}. 

\subsubsection{Consistency Score}
The integrity constraints vary depending on the domain and the application conditions. Therefore, they are defined in problem-specific settings. For example, in some domains, data observations cannot be negative or their values should fall within a particular range. After defining the integrity rules for the data values, the consistency score can be calculated as the fraction of data values that do not meet the integrity constraints \cite{pipino2002data}: $Consistency=\frac{NCV}{N}$, where $NCV$ is the number of consistent values and $N$ is the size of the data window. Data consistency is a fundamental issue for data quality, as it checks data conflicts that can detect errors in the data recording process \cite{fan2015data}.

%%% Bestoun 11 January 2023

\subsubsection{Timeliness Score}
Timeliness, or data currency, is a semantic measure that characterizes whether the data fits the application domain \cite{johnson2016application}. To quantify the timeliness of the data, we used a goodness-of-fit test. Goodness-of-fit tests measure the likelihood that current data are sampled from a specific cumulative distribution function(cdf) or probability density function (pdf) of the underlying data-generating distribution \cite{aslam2021new}. Several goodness-of-fit test techniques can be applied according to the nature of the data \cite{d2017goodness}. The most popular tests are the Kolmogorov-Smirnov, Anderson-Darling, and Cramér-von Mises statistical tests \cite{evans2008distribution}. Each test calculates test statics that is interpreted for the fitness of the data with the compared distribution.

In our experiments, a two-sample Kolmogorov-Smirnov statistical test is used to calculate the goodness-of-fit metric. Kolmogorov-Smirnov test is a non-parametric statistical tool to determine whether two samples are drawn from the same distribution \cite{pratt2012concepts}. The main motivation for adopting the Kolmogorov-Smirnov test is that it is a powerful method for small subsets \cite{augste2011relative}, as in our use-case settings. For two empirical cumulative distribution functions $\hat{F_1}$ and $\hat{F_2}$ for two independent
random samples $X=X_1, \ldots, X_n$ and $Y=Y_1, \ldots, Y_m$ respectively,
the Kolmogorov–Smirnov test statistic is computed as \cite{kvam2022nonparametric}:
\begin{equation}
KS=\max _{1 \leqslant i \leqslant N}\left|\hat{F_1}\left(Z_i\right)-\hat{F_2}\left(Z_i\right)\right|,
\end{equation}
where $Z$ is the combined sample of $X$ and $Y$, $N=n+m$.

\subsubsection{Skewness Score.}
In real-world applications, especially the Internet of Things (IoT), where data are collected from sensors, distributional drift (or shift) is one of the most frequent data issues in the system \cite{foidl2019risk}. Data flows are validated against distributional deviations that induce skewness in the data distribution \cite{caveness2020tensorflow}. To calculate the distributional skewness score, the divergence magnitude can be calculated to measure the dissimilarity between the distributions of the current data window and the historical data \cite{lee2000measures}. There are numerous methods to calculate the divergence measure; Jensen-Shannon (JSD) and Kullback-Leibler (KLD) Divergences are the most popular ones \cite{nguyen2015non}.

For our framework, we have used the JSD metric to calculate the skewness score. JSD metric is a symmetrization of the KLD metric. The main property of JSD is that it is bounded in the interval $[0,1]$, while the KLD value may be infinite \cite{lionis2021rssi}. According to JSD, the dissimilarity between two probability distributions $P$ and $Q$ is calculated as \cite{lin1991divergence}:
\begin{equation}
\label{entropy}
\small\mathrm{JSD}(P\| Q) = \mathrm{JSD}(P\| Q) :=
H\left(\frac{P+Q}{2}\right)-\frac{H(P)+H(Q}{2},
\end{equation}
where the function $H$ denotes Shannon's entropy and is given by:
\begin{equation}
H(p)=-\int p(\mathbf{Y}) \log p(\mathbf{Y}) d \mathbf{Y}.
\end{equation}

%%% Bestoun 11 January 2023

A final remark on the data quality dimensions defined in this section is that other definitions can be introduced according to the task. For example, accuracy, completeness, and consistency scores can be defined as ‘1’ if the data window includes the corresponding data quality issue and ‘0’ otherwise \cite{blake2011effects}, rather than using the fraction of erroneous data instances. This can be adopted in safety-critical domains with a lower tolerance level for faulty data, such as medical applications and autonomous cars \cite{chan2022fault}. In our industrial use cases, using fractions of erroneous data is more suitable since potentially dropping data instances (i.e., giving them a score of ‘1’) is not preferred.

\subsection{Finding the Consolidated Data Quality Score}
After calculating the score values of the data quality dimensions, the values are aggregated in a consolidated score that represents the quality of the data window \cite{cichy2019overview}. The most basic approach to aggregate the metrics is by taking the arithmetic mean, or its variants, such as the weighted average, of the data quality dimensions. However, it was argued that the arithmetic mean does not provide sound aggregation \cite{heinrich2018requirements}. Additionally, aggregation methods are sensitive to the scales of the variables. Therefore, the calculation of the aggregated metric will be dominated by the dimensions with large scales, and the effect of the dimensions with low scales will be negligible. 

Before aggregating the quality scores, the matrix $DQ_{nm}$ that contains the quality scores of the $n$ data instance of $m$ dimension is standardized. The standardization step is essential before aggregation so that the different data quality metrics can be equally integrated into the overall quality score \cite{decancq2012inequality}. The standardization, also known as whitening, can be achieved by calculating the z-score of each element $dq_{ij}$ of the matrix $DQ_{nm}$: $z_{ij}=\frac{(dq_{ij}-\mu_{j})}{\sigma_{j}}$, where $\mu_{j}$ is the mean $\mu_{j}$ is the standard deviation of the respective column. The standardization process would give the different data quality dimensions uniform weights when calculating the aggregation metric. 

In a recent survey \cite{teh2020sensor}, it was shown that the principal component analysis-based methods are the most widely used techniques to detect data errors. Therefore, we follow the literature and use principal component analysis (PCA) to extract the score that represents the overall quality of the data windows. PCA is a widely used dimensionality reduction technique that extracts the most crucial information from the data. PCA maps each data instance of $d$ dimensional space to $k$ dimensional space using an orthogonal transformation such that $k<d$. The first principal component is constructed by finding the direction of maximum variance \cite{abdi2010principal}. The principal components of dataset $X$ are found by solving the eigenvalue problem:
\begin{equation}
\Sigma \mathbf{M}=\lambda \mathbf{M},
\end{equation}
where $\Sigma$ is the covariance matrix $\sum=XX^{T}$ that measures the correlation between the variables of the original data $X$, $\lambda$ is the vector of eigenvalues, and $M$ is the matrix of eigenvectors that contains the principal components ordered by the size of their eigenvalue.

%%% Bestoun 11 January 2023

\subsection{Initialization Phase}
\label{sec:init}
Before deployment, the DQSOps framework requires preparing a warm-start of the ML predictor using training data that includes ground-truth quality scores. The initialization phase of DQSOps is indicated in Figure \ref{fig:DSOps} by the red path. Through this phase, a reliable ML predictor will be delivered to be put into practice in production. To produce the ML predictor, ground-truth labels are found, as discussed in the previous sections, and stored in a repository that will be used to train the ML model. The prediction accuracy of the ML model is monitored until it reaches a predetermined threshold $\tau$ evaluated with respect to a performance metric. The ML predictor is deployed in the real-world problem upon reaching this threshold. In this phase, the mutant simulator component plays an essential role in the learning task. The mutants would help the model learn to identify data quality issues by providing the model with more extensive and diverse quality issues that could potentially arise in real-life environments. Furthermore, the component accelerates the learning process, compared to waiting for data quality issues that may not be frequent in reality, which can be time-consuming. 

In addition to producing the ML predictor, other meta-information files are prepared in this phase to be used in practice. These files are the anomaly detection model used to calculate the accuracy score and a reference data distribution used to calculate the skewness score. These files should be prepared using high-quality clean data to enhance the reliability of the ground-truth quality scores. The paths of these files are then stored in the configuration file to be loaded once the DQSOps framework is eventually deployed in production.

%%% Bestoun 11 January 2023

\subsection{Method Activator Component}
\label{sec:activ}
\begin{figure}
    \centering
    \includegraphics[scale=0.5]{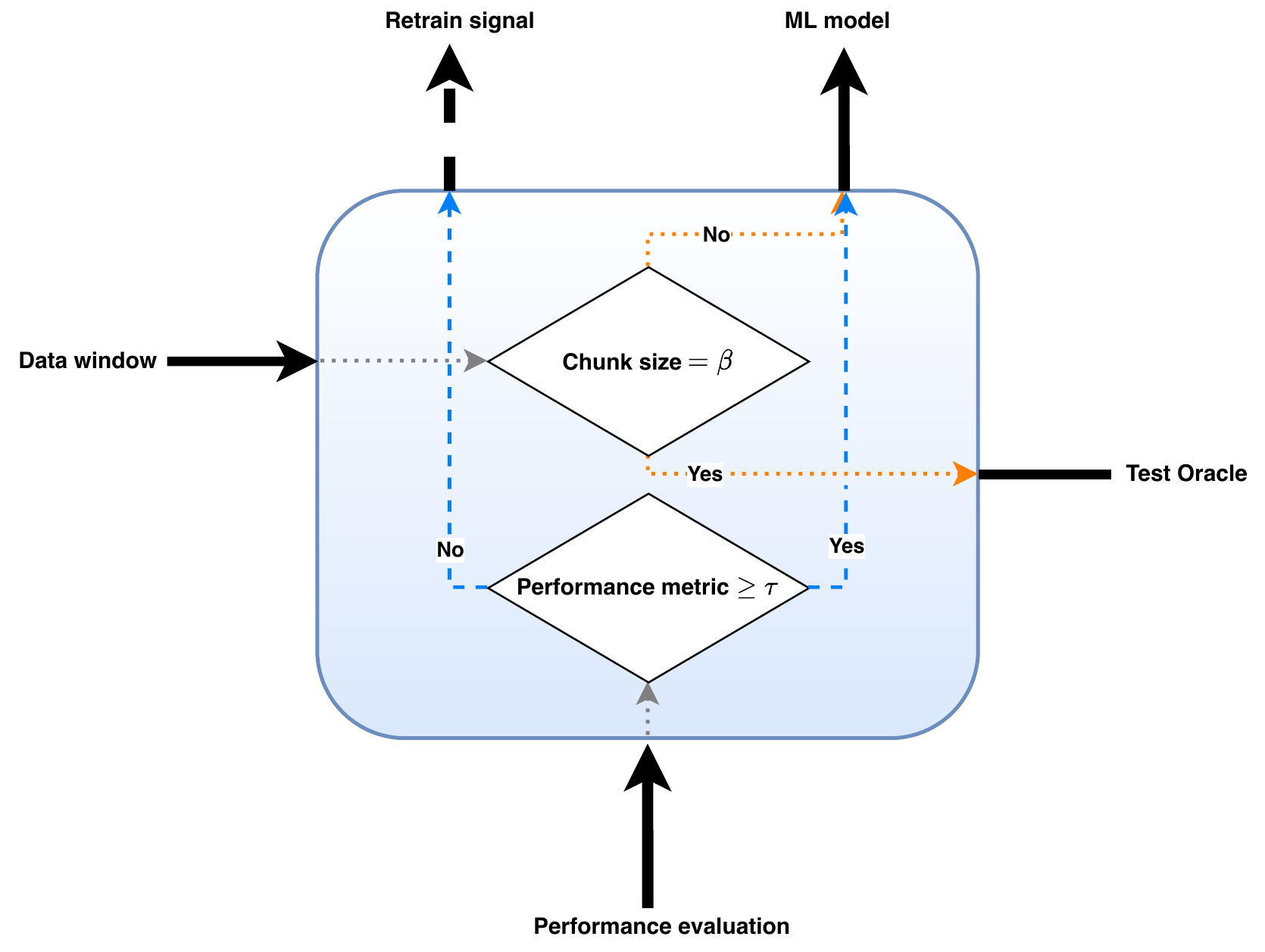}
    \caption{Flowchart diagram of the method activator component}
    \label{fig:activ}
\end{figure}

\begin{figure}
    \centering
    \includegraphics[scale=0.81]{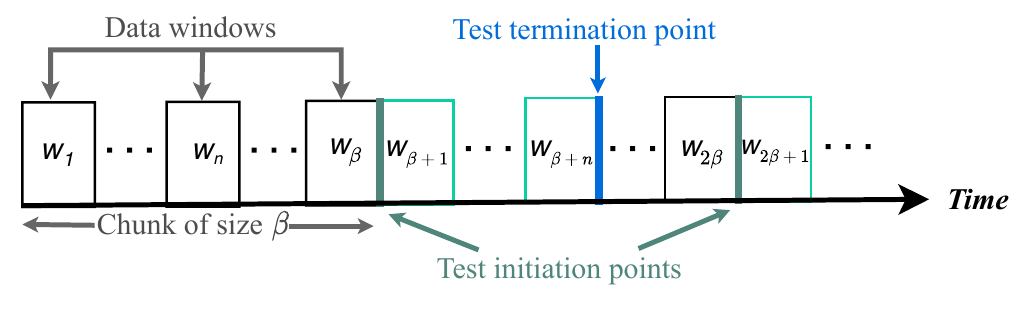}
    \caption{Mechanism of processing the data streams}
    \label{fig:blocks}
\end{figure}

One of the core elements of DQSOps is the method activator which orchestrates the pipeline flow. As illustrated in the decision flowchart in Figure \ref{fig:activ}, the component uses a set of predetermined criteria to execute the appropriate approach for data quality scoring. Once the data window is collected, a decision is made based on the chunk size. The activator repeatedly executes the ML model to obtain the data quality scores until the chunk size reaches a preconfigured threshold $\beta$. Figure \ref{fig:blocks} illustrates the mechanism for processing the data windows and chunks. The threshold $\beta$ represents how frequently we want to test the ML model performance. Once the threshold is reached by collecting $\beta$ data windows, the activator executes an evaluation methodology using the standard-based approach. Subsequently, the standard approach produces the ground-truth scores of $n$ data windows, where $1\leq n<\beta$. This is necessary to perform continuous monitoring of the ML model performance. 

% The prediction-based approach is only meaningful when the number of data windows $n$ is less than the block size $\beta$. When the number of data windows is equal to the block size, the prediction-based approach becomes unnecessary and the standard-based would be sufficient to score the data windows. However, the main advantage of using such prediction-based approach instead of going through the standard scoring is that, in practice, standard-based approach may not be able to score all data windows due to its high cost that prevent it from processing rapid data streams.

The evaluation is conducted using a specified test oracle \cite{li2016test}, as shown in Figure \ref{fig:DSOps}. The specified test oracle requires oracle information and oracle procedure \cite{barr2014oracle}. The oracle information characterizes the expected output, which is obtained by the ground-truth scores by the standard-based approach. For the oracle procedure, a relevant performance metric, such as the prediction error, is used to evaluate the predicted quality score of the ML model using the oracle information. Afterward, the result of the test oracle is forwarded to the activator, which compares it with a tolerance level $\tau$. The level represents the minimum accuracy required to achieve and can be set based on quality requirements or inferred by observing the ML performance in the initialization phase; see Section \ref{sec:init}. If the performance of the ML model falls below the desired level, a retrain signal is sent to update the ML model using the newly obtained data quality scores. This strategy ensures the efficiency and reliability of the ML model over time.

\section{Implementation and Evaluation}
\label{sec:results}
The framework is applied and evaluated in a real-world industrial use case. The production results are analyzed in terms of predictive performance and time-efficiency criteria. An analysis of the effects of the mutation percentage on the initialization phase is also presented.

\subsection{Industrial Use Case Description}
The industrial use case in which we implement our framework is the industrial process of Electroslag Remelting (ESR) vacuum pumping at Uddeholm steel company in Sweden\footnote{https://www.uddeholm.com/}. The vacuum pump is used to ensure the production of high-quality steel. This is achieved by extracting the air oxygen from the furnace. In production, each vacuum pump event lasts up to 20 minutes. The observation of interest is the pressure value generated inside the vacuum chamber. To record the pressure values, a sensor is connected to the furnace and registers the value every millisecond. The pressure data windows are continuously transmitted every second through Apache Kafka streaming platform\footnote{https://kafka.apache.org/} to enable real-time analysis. The pressure value should gradually decrease in a proper pump event until it reaches the desired minimum value within 20 minutes. However, in some cases, an improper pump event occurs, and the minimum pressure value is not met so the furnace will be re-initiated. Therefore, it is essential to distinguish the status of the pump events. The task is to quantify the quality of the pressure data collected from each pump event using the proposed DQSOps framework. For a production environment, DQSOps is developed using Python programming language and deployed in the system through a Docker container\footnote{https://www.docker.com/}. The Docker ecosystem allows the deployment flexibly and efficiently regardless of the underlying system. For ML models, we use two popular decision tree-based algorithms, random forests (RF) \cite{breiman2001random} and extreme gradient boosting (XGBoost) \cite{chen2016xgboost} models. RF and XGBoost are two efficient ensemble algorithms that have been widely used in various applications due to their simplicity, high performance, and interpretability property, making them suitable for industrial applications \cite{kiangala2021effective}.

%%% Bestoun 11 January 2023

\subsection{Predictive Performance Evaluation}
The performance results of the prediction-based methods of our framework are evaluated in the production environment of our use case. These methods are based on the regression algorithm that predicts the data quality score rather than calculating it using the standard approach. The performance is evaluated using two performance metrics, the mean absolute error (MAE) and the coefficient of determination $R^2$ given by the formulae:
\begin{equation}
\text {MAE}=\frac{\sum_{i=1}^n\left|y_i-\hat{y}_i\right|}{n},
\end{equation}
\begin{equation}
R^2 = 1-\frac{\sum_{i=1}^n\left(y_i-\hat{y}_i\right)^2}{\sum_{i=1}^n\left(y_i-\bar{y}\right)^2},
\end{equation}
where $y_i$ is the ground-truth score calculated using the oracle-based approach, $\hat{y}_i$ is the predicted value using the regression algorithm, and $\bar{y}$ represents the mean of all scores.

Table \ref{tab:rf_perf} shows the performance of the RF prediction-based method, and Table \ref{tab:xgb_perf} shows the performance of the XGBoost prediction-based method. The arithmetic mean of the error rates of all possible combinations of data quality dimensions of the same size is recorded. We have also calculated the Coefficient of Variation (CV) to measure the relative dispersion of the error rates of each experiment. CV is calculated as the ratio between the standard deviation (std) $\sigma$ and the mean $\mu$ of the population: $CV=\frac{\sigma}{\mu}$. Moreover, to provide insights into the distribution of the DQ scores and to assess the significance of the error rates compared to the statistics of the DQ scores, we computed the summary statistics for the DQ score values presented in Table \ref{tab:dqs_stats} for the different experimental settings. These statistics offer an overview of the DQ score ranges and their standard deviation, which allows us to contextualize the performance of the ML models. 

As for the predictive performance, for both ML algorithms, the MAE value increases with the number of quality dimensions, as the true scoring function becomes more difficult to capture by the algorithms. The MAE metric's average value is low in experiments that are performed using a single data quality dimension. This is because most of the data windows are of high quality, so their quality score is close to 0. However, the $R^2$ metric increases with the size of the data quality dimensions for both RF and XGBosst as the algorithms fit the data better. From the tables, we can also see that the difference in performance is not significant between RF and XGBoost, with the latter showing slightly lower MAE and higher $R^2$ values. However, the significant run-time efficiency achieved by the XGBoost model, as will be presented in the subsequent analysis, makes it the most superior ML model to deploy in the DQSOps framework. 

\begin{table}
\centering
\caption{Average performance of the RF prediction-based method}
\begin{tabular}{c|lll|lll} 
\toprule
\multirow{2}{*}{\textbf{DQD}} & \multicolumn{3}{c|}{\textbf{MAE}}           & \multicolumn{3}{c}{\textbf{$\mathrm{R^2}$}}             \\ 
\cline{2-7}
                              & \textbf{mean} & \textbf{std} & \textbf{CV} & \textbf{mean} & \textbf{std} & \textbf{CV}  \\ 
\midrule
\textbf{1}                    & 0.0148        & 0.0052       & 0.3529       & 0.7133        & 0.2111       & 0.2959        \\ 

\textbf{2}                    & 0.4142        & 0.0561       & 0.1353       & 0.7325        & 0.2289       & 0.3125        \\ 

\textbf{3}                    & 0.5604        & 0.0928       & 0.1656       & 0.7978        & 0.1353       & 0.1696        \\ 

\textbf{4}                    & 0.6411        & 0.1032       & 0.1610       & 0.8435        & 0.0362       & 0.0430        \\ 

\textbf{5}                    & 0.6132        & 0.0000       & 0.0000       & 0.8936        & 0.0000       & 0.0000        \\
\bottomrule
\end{tabular}
\label{tab:rf_perf}
\end{table}

\begin{table}
\centering
\caption{Average performance of the XGBoost prediction-based method}
\begin{tabular}{c|lll|lll} 
\toprule
\multirow{2}{*}{\textbf{DQD}} & \multicolumn{3}{c|}{\textbf{MAE}}           & \multicolumn{3}{c}{\textbf{$\mathrm{R^2}$}}             \\ 
\cline{2-7}
                              & \textbf{mean} & \textbf{std} & \textbf{CV} & \textbf{mean} & \textbf{std} & \textbf{CV}  \\ 
\midrule
\textbf{1}                    & 0.0138        & 0.0036       & 0.2617       & 0.7820        & 0.1690       & 0.2161        \\
\textbf{2}                    & 0.4242        & 0.0619       & 0.1459       & 0.7231        & 0.2341       & 0.3237        \\
\textbf{3}                    & 0.5483        & 0.1044       & 0.1904       & 0.7846        & 0.1919       & 0.2445        \\
\textbf{4}                    & 0.5934        & 0.1075       & 0.1812       & 0.8582        & 0.0641       & 0.0746        \\
\textbf{5}                    & 0.6320        & 0.0000       & 0.0000       & 0.8916        & 0.0000       & 0.0000        \\
\bottomrule
\end{tabular}
\label{tab:xgb_perf}
\end{table}

\begin{table}
\centering
\caption{Statistics of data quality scores}
\begin{tabular}{c|c|c}
\toprule
\textbf{DQD} & \textbf{DQ score range} & \textbf{DQ scores std} \\
\midrule
\textbf{1} & [0, 0.9345] & 0.0896 \\
\textbf{2} & [-15.173, 4.820] & 2.27 \\
\textbf{3} & [-13.819, 5.472] & 2.355 \\
\textbf{4} & [-12.204, 8.152] & 2.93 \\
\textbf{5} & [-9.899, 9.855] & 2.846 \\
\bottomrule
\end{tabular}
\label{tab:dqs_stats}
\end{table}

\subsection{Time Speedup Evaluation}
To evaluate the run-time efficiency, we compare the time required to execute the standard and prediction-based data quality scoring approaches. Experiments were conducted with different numbers of data quality dimensions to analyze the effect of the size of the data quality dimensions on the execution time. For example, we have five possible combinations in total for a single data quality dimension. Each combination evaluates an individual data quality dimension separately. While for two data quality dimensions, ten pairs of data quality dimensions can be constructed from the five overall quality dimensions defined in our framework, and so on. Table~\ref{tab:seconds} summarizes the average computation time of each method per data window in seconds.

\begin{table*}
\centering
\caption{Time required to score a data window using the different approaches in seconds}
\label{tab:seconds}
\begin{tabular}{c|lll|lll|c|lll|c} 
\toprule
\multirow{2}{*}{\textbf{DQD}} & \multicolumn{3}{c|}{\textbf{Standard-based}}        & \multicolumn{4}{c|}{\textbf{RF prediction-based}}                                        & \multicolumn{4}{c}{\textbf{XGBoost prediction-based}}                                                         \\ 
\cline{2-12}
                              & \textbf{mean} & \textbf{std} & \textbf{CV} & \textbf{mean} & \textbf{std} & \textbf{CV} & \multicolumn{1}{l|}{\textbf{Speedup}} & \textbf{mean} & \textbf{std} & \textbf{CV} & \multicolumn{1}{l}{\textbf{\textbf{Speedup}}}  \\ 
\midrule
\textbf{1}                    & 0.19057       & 0.32218      & 1.69066      & 0.01953       & 0.00142      & 0.07260      & \textbf{9.76x}                        & 0.00050       & 0.00004      & 0.08774      & \textbf{381.90x}                               \\
\textbf{2}                    & 0.37671       & 0.39409      & 1.04614      & 0.01268       & 0.00035      & 0.02740      & \textbf{29.71x}                       & 0.00103       & 0.00007      & 0.06453      & \textbf{366.45x}                               \\
\textbf{3}                    & 0.57191       & 0.40000      & 0.69941      & 0.01259       & 0.00042      & 0.03334      & \textbf{45.44x}                       & 0.00109       & 0.00004      & 0.03334      & \textbf{523.72x}                               \\
\textbf{4}                    & 0.75229       & 0.32335      & 0.42983      & 0.01245       & 0.00030      & 0.02405      & \textbf{60.42x}                       & 0.00118       & 0.00004      & 0.03472      & \textbf{638.07x}                               \\
\textbf{5}                    & 0.94101       & 0.00000      & 0.00000      & 0.01265       & 0.00000      & 0.00000      & \textbf{74.38x}                       & 0.00115       & 0.00000      & 0.00000      & \textbf{818.98x}                               \\
\bottomrule
\end{tabular}
\end{table*}

Analogously to the analysis presented in the previous section, we analyzed the computational run-time of different sizes of data quality dimensions. We calculated the summary statistics and Coefficient of Variation (CV) for the computational run-time of each experiment to explore the dispersion of the results. We can observe that the CV is higher for the standard-based scoring method. This is because calculating completeness and consistency scores requires less computational time than calculating the goodness-of-fit or skewness. A simple fraction is required to calculate completeness and consistency scores, while goodness-of-fit and skewness involve estimating probability distributions. Hence, variability is higher when conducting experiments with a single data quality dimension. The CV values decrease as the number of data quality dimensions increases, and the run-time variation of the different experiments decreases. However, the standard deviation values are lower, and the CV values are closer for prediction-based scoring methods than for the standard-based approach. This signifies that the level of dispersion around the mean is lower, and the execution times are relatively closer for prediction-based methods.

For the standard-based method, the average computational time is proportional to the number of quality dimensions. In contrast, the average computational time does not depend on the number of quality dimensions for prediction-based approaches. We can see that the summary statistics are similar for each ML model used to predict the data quality score in all experiments. This finding has substantial practical implications on the sensitivity to the number of quality dimensions for production environments. The run-time for the prediction-based scoring process is agnostic to the number of quality dimensions. In contrast, the run-time would increase with the number of quality dimensions for the standard scoring process. Regarding speedup rates, the results in Table \ref{tab:seconds} show significant levels of run-time efficiency with the prediction-based quality scoring approach. Specifically, the random forest (RF) prediction-based method registers a 10x speedup increase over the standard-based method for a single quality dimension. It reaches approximately 75x when using all the data quality dimensions. Similarly, for the XGBoost prediction-based method, the speedup is approximately 382x when using one quality dimension to roughly reach 819x when using all quality dimensions. However, comparing the different ML models, we see that the XGBoost algorithm requires significantly less run-time than RF. This is because RF relied on more complex ensembles of decision trees to make predictions than XGBoost during the hyperparameter tuning phase.

% \begin{figure*}

% \begin{subfigure}[b]{0.5\textwidth} 
%     \centering
%     \includegraphics[scale=0.62]{Figures/MAE_VP.pdf}
%     \caption{MAE}
%     \label{fig:MAE}
% \end{subfigure}%
% \begin{subfigure}[b]{0.5\textwidth} 
%     \centering
%     \includegraphics[scale=0.62]{Figures/R2.pdf}
%     \caption{R2}
%     \label{fig:R2}
% \end{subfigure}%
% \caption{Performance metrics}
%     \label{fig:perfromance_metrics}
% \end{figure*}

\subsection{Mutation Percentage Impact}
\begin{figure*}
\begin{subfigure}[b]{0.5\textwidth} 
    \centering
    \includegraphics[scale=0.215]{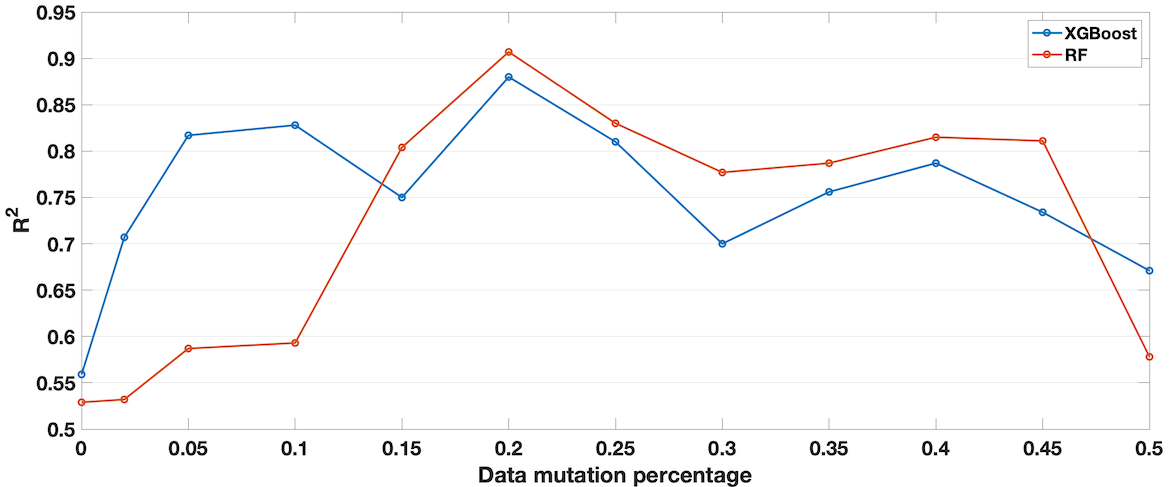}
    \caption{R2}
    \label{fig:perc_R2}
\end{subfigure}%
\hfill
\begin{subfigure}[b]{0.5\textwidth}
    \centering
    \includegraphics[scale=0.215]{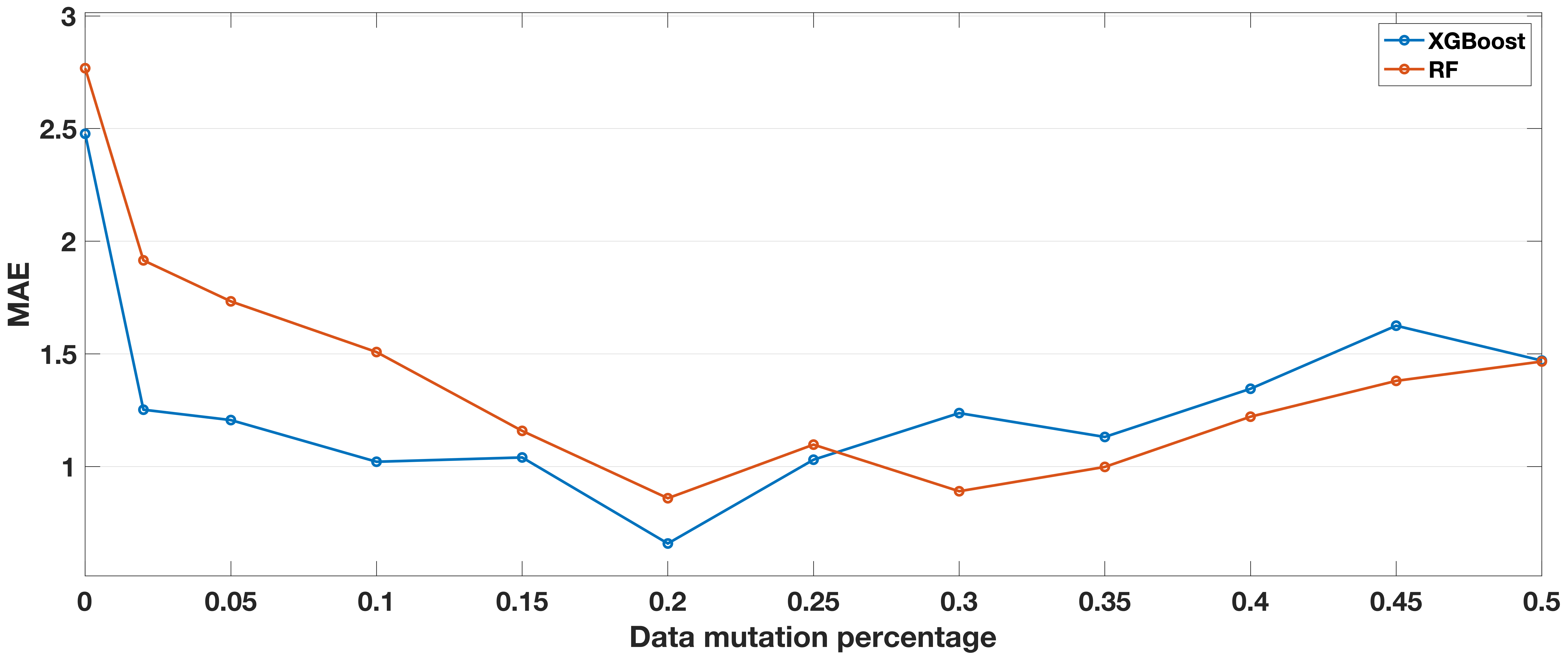}
    \caption{MAE}
    \label{fig:per_mae}
\end{subfigure}
    \caption{Mutants percentage impact on performance}
    \label{fig:mut_perc}
\end{figure*}

As discussed in Section \ref{sec:methodology}, data mutation methodology was carried out to simulate issues that affect the quality of the processed data window according to a specified percentage of mutations. Specifically, we utilized the whole set of data quality dimensions, which are five in total, in this experiment. To analyze the impact of the mutation percentage in the initialization phase, we have observed the result of the test oracle by adjusting the percentage value using a validation set. The test oracle was evaluated in terms of $R^2$ and MAE performance metrics for each ML algorithm. The results of the RF and XGBoost algorithms are summarized in Figure \ref{fig:mut_perc}. 

For both algorithms, we can see that the evaluation metrics demonstrate a U-shaped trend. Specifically, for $R^2$ as displayed in Figure \ref{fig:perc_R2}, the value begins at a rate slightly above 0.5 for both algorithms when no data mutants are simulated, then improving to reach approximately 0.9 using a data mutation percentage of 20\%. After reaching the peak, the performance metric starts to drop as we increase the percentage of data mutants to reach approximately 0.68 and 0.58 for XGBoost and RF, respectively, at a mutation percentage of 50\%. Identically as shown in Figure \ref{fig:per_mae}, the curve of the MAE metric follows the same trend. Beginning with the highest error levels with noise mutants introduced in the training data to achieve optimal performance with MAE of 0.65 for XGBoost and 0.86 for RF at 20\% of the percentage of mutants. After that, inducing more data mutants does not improve the training process and the error steadily increases with higher mutation percentages.

The results of this experiment showed that the percentage of induced mutants should be carefully chosen to obtain the value that produces the optimal performance. In particular, a low mutation percentage may not introduce enough data quality issues to sufficiently train the ML model to correctly score the data quality. However, setting the mutation percentage too high could introduce many quality issues into the data, disrupting the learning process and leading to poor performance. Therefore, choosing the mutation percentage carefully to strike a balance between introducing sufficient data quality issues and avoiding introducing too many errors that could negatively impact the ML performance is critical in the initialization phase.

% \begin{figure*}
%     \centering
%     \includegraphics[scale=0.4]{Figures/R2_results.png}
%     \caption{R2}
%     \label{fig:r2}
% \end{figure*}

% \begin{figure*}
%     \centering
%     \includegraphics[scale=0.4]{Figures/R2_results.png}
%     \caption{MAE}
%     \label{fig:mae}
% \end{figure*}

\section{Threats to External Validity}
\label{sec:threats}
External validity concerns generalizing the findings of the proposed framework. Generalizability is the main threat in constructing frameworks for real-world applications since every use case may require a different approach to solve the problem. However, the DQSOps framework shown in Figure \ref{fig:DSOps} presents an abstract pipeline workflow and was designed to be flexible without rigid specifications on its components to promote reproducibility. Therefore, it can be employed for scoring data of different natures in diverse settings with simple adjustments to the methods used in this research. For example, the scores of data quality dimensions presented in Section \ref{sec:scoring} were designed to handle univariate time-series data as our use case requires. However, other methods can be followed to score the quality of multivariate data \cite{taleb2021big, martens2001multivariate} and can be integrated into the DQSOps framework. Similarly, the rest of the components can be determined to fit the system requirements of the handled use case. Additionally, the utilization of the configuration files enhanced delivering a flexible framework. The configuration files store auxiliary meta-information such as the parameters of data quality dimensions, and mutation percentage. This meta-information can be configured based on the use case needs without modifying the underlying structure of DQSOps. Based on our analysis, we believe that using different techniques than those employed in this research would still yield similar results and findings to our study.

\section{Conclusion}
\label{sec:conclusion}
In this paper, we have presented a Data Quality Scoring Operations (DQSOps) framework. The framework can quantify the quality of data records and distinguish high- and low-quality data. The framework is integrated with two scoring approaches: a prediction-based approach and a standard-based approach. The prediction-based approach is used to predict the quality score of the collected data windows using an ML model. In contrast, the standard-based approach is periodically invoked to produce the ground-truth quality score. The score is found by combining several score metrics from the defined data quality dimensions and is used to design a test oracle. The test oracle continuously evaluates the ML model to activate a retrain signal to update the ML model. Furthermore, a data mutants simulator is integrated into DQSOps to induce quality issues in the data and facilitate the learning process. The framework is deployed and evaluated in a real-world industrial use case. The results showed significant speedup rates achieved by DQSOps compared to the standard scoring method while maintaining high predictive performance. An analysis of the optimal mutation percentage was also presented to gain insight into its impact on the learning process.

\begin{acks}
This work has been funded by the Knowledge Foundation of Sweden (KKS) through the Synergy Project AIDA - A Holistic AI-driven Networking and Processing Framework for Industrial IoT (Rek:20200067).
\end{acks}
% \begin{acks}
% \end{acks}

\balance
\bibliographystyle{ACM-Reference-Format}
\bibliography{sample-base}


\begin{thebibliography}{56}

%%% the Web address.

\ifx \showCODEN    \undefined \def \showCODEN     #1{\unskip}     \fi
\ifx \showDOI      \undefined \def \showDOI       #1{#1}\fi
\ifx \showISBNx    \undefined \def \showISBNx     #1{\unskip}     \fi
\ifx \showISBNxiii \undefined \def \showISBNxiii  #1{\unskip}     \fi
\ifx \showISSN     \undefined \def \showISSN      #1{\unskip}     \fi
\ifx \showLCCN     \undefined \def \showLCCN      #1{\unskip}     \fi
\ifx \shownote     \undefined \def \shownote      #1{#1}          \fi
\ifx \showarticletitle \undefined \def \showarticletitle #1{#1}   \fi
\ifx \showURL      \undefined \def \showURL       {\relax}        \fi
% invisible to TeX
\providecommand\bibfield[2]{#2}
\providecommand\bibinfo[2]{#2}
\providecommand\natexlab[1]{#1}
\providecommand\showeprint[2][]{arXiv:#2}

\bibitem[Abdi and Williams(2010)]%
        {abdi2010principal}
\bibfield{author}{\bibinfo{person}{Herv{\'e} Abdi} {and}
  \bibinfo{person}{Lynne~J Williams}.} \bibinfo{year}{2010}\natexlab{}.
\newblock \showarticletitle{Principal component analysis}.
\newblock \bibinfo{journal}{\emph{Wiley interdisciplinary reviews:
  computational statistics}} \bibinfo{volume}{2}, \bibinfo{number}{4}
  (\bibinfo{year}{2010}), \bibinfo{pages}{433--459}.
\newblock


\bibitem[Aslam(2021)]%
        {aslam2021new}
\bibfield{author}{\bibinfo{person}{Muhammad Aslam}.}
  \bibinfo{year}{2021}\natexlab{}.
\newblock \showarticletitle{A new goodness of fit test in the presence of
  uncertain parameters}.
\newblock \bibinfo{journal}{\emph{Complex \& Intelligent Systems}}
  \bibinfo{volume}{7}, \bibinfo{number}{1} (\bibinfo{year}{2021}),
  \bibinfo{pages}{359--365}.
\newblock


\bibitem[Augste and Lames(2011)]%
        {augste2011relative}
\bibfield{author}{\bibinfo{person}{Claudia Augste} {and}
  \bibinfo{person}{Martin Lames}.} \bibinfo{year}{2011}\natexlab{}.
\newblock \showarticletitle{The relative age effect and success in German elite
  U-17 soccer teams}.
\newblock \bibinfo{journal}{\emph{Journal of sports sciences}}
  \bibinfo{volume}{29}, \bibinfo{number}{9} (\bibinfo{year}{2011}),
  \bibinfo{pages}{983--987}.
\newblock


\bibitem[Barr et~al\mbox{.}(2014)]%
        {barr2014oracle}
\bibfield{author}{\bibinfo{person}{Earl~T Barr}, \bibinfo{person}{Mark Harman},
  \bibinfo{person}{Phil McMinn}, \bibinfo{person}{Muzammil Shahbaz}, {and}
  \bibinfo{person}{Shin Yoo}.} \bibinfo{year}{2014}\natexlab{}.
\newblock \showarticletitle{The oracle problem in software testing: A survey}.
\newblock \bibinfo{journal}{\emph{IEEE transactions on software engineering}}
  \bibinfo{volume}{41}, \bibinfo{number}{5} (\bibinfo{year}{2014}),
  \bibinfo{pages}{507--525}.
\newblock


\bibitem[Batini et~al\mbox{.}(2015)]%
        {batini2015data}
\bibfield{author}{\bibinfo{person}{Carlo Batini}, \bibinfo{person}{Anisa Rula},
  \bibinfo{person}{Monica Scannapieco}, {and} \bibinfo{person}{Gianluigi
  Viscusi}.} \bibinfo{year}{2015}\natexlab{}.
\newblock \showarticletitle{From data quality to big data quality}.
\newblock \bibinfo{journal}{\emph{Journal of Database Management (JDM)}}
  \bibinfo{volume}{26}, \bibinfo{number}{1} (\bibinfo{year}{2015}),
  \bibinfo{pages}{60--82}.
\newblock


\bibitem[Blake and Mangiameli(2011)]%
        {blake2011effects}
\bibfield{author}{\bibinfo{person}{Roger Blake} {and} \bibinfo{person}{Paul
  Mangiameli}.} \bibinfo{year}{2011}\natexlab{}.
\newblock \showarticletitle{The effects and interactions of data quality and
  problem complexity on classification}.
\newblock \bibinfo{journal}{\emph{Journal of Data and Information Quality
  (JDIQ)}} \bibinfo{volume}{2}, \bibinfo{number}{2} (\bibinfo{year}{2011}),
  \bibinfo{pages}{1--28}.
\newblock


\bibitem[Bosch et~al\mbox{.}(2021)]%
        {bosch2021engineering}
\bibfield{author}{\bibinfo{person}{Jan Bosch},
  \bibinfo{person}{Helena~Holmstr{\"o}m Olsson}, {and} \bibinfo{person}{Ivica
  Crnkovic}.} \bibinfo{year}{2021}\natexlab{}.
\newblock \showarticletitle{Engineering ai systems: A research agenda}.
\newblock \bibinfo{journal}{\emph{Artificial Intelligence Paradigms for Smart
  Cyber-Physical Systems}} (\bibinfo{year}{2021}), \bibinfo{pages}{1--19}.
\newblock


\bibitem[Breiman(2001)]%
        {breiman2001random}
\bibfield{author}{\bibinfo{person}{Leo Breiman}.}
  \bibinfo{year}{2001}\natexlab{}.
\newblock \showarticletitle{Random forests}.
\newblock \bibinfo{journal}{\emph{Machine learning}} \bibinfo{volume}{45},
  \bibinfo{number}{1} (\bibinfo{year}{2001}), \bibinfo{pages}{5--32}.
\newblock


\bibitem[Byabazaire et~al\mbox{.}(2022)]%
        {byabazaire2022end}
\bibfield{author}{\bibinfo{person}{John Byabazaire}, \bibinfo{person}{Gregory
  O’Hare}, {and} \bibinfo{person}{Declan~T Delaney}.}
  \bibinfo{year}{2022}\natexlab{}.
\newblock \showarticletitle{End-to-End Data Quality Assessment Using Trust for
  Data Shared IoT Deployments}.
\newblock \bibinfo{journal}{\emph{IEEE Sensors Journal}}
  (\bibinfo{year}{2022}).
\newblock


\bibitem[Cappiello et~al\mbox{.}(2018)]%
        {cappiello2018validating}
\bibfield{author}{\bibinfo{person}{Cinzia Cappiello}, \bibinfo{person}{C
  Cerletti}, \bibinfo{person}{C Fratto}, {and} \bibinfo{person}{Barbara
  Pernici}.} \bibinfo{year}{2018}\natexlab{}.
\newblock \showarticletitle{Validating data quality actions in scoring
  processes}.
\newblock \bibinfo{journal}{\emph{Journal of Data and Information Quality
  (JDIQ)}} \bibinfo{volume}{9}, \bibinfo{number}{2} (\bibinfo{year}{2018}),
  \bibinfo{pages}{1--27}.
\newblock


\bibitem[Caveness et~al\mbox{.}(2020)]%
        {caveness2020tensorflow}
\bibfield{author}{\bibinfo{person}{Emily Caveness},
  \bibinfo{person}{Paul~Suganthan GC}, \bibinfo{person}{Zhuo Peng},
  \bibinfo{person}{Neoklis Polyzotis}, \bibinfo{person}{Sudip Roy}, {and}
  \bibinfo{person}{Martin Zinkevich}.} \bibinfo{year}{2020}\natexlab{}.
\newblock \showarticletitle{Tensorflow data validation: Data analysis and
  validation in continuous ml pipelines}. In
  \bibinfo{booktitle}{\emph{Proceedings of the 2020 ACM SIGMOD International
  Conference on Management of Data}}. \bibinfo{pages}{2793--2796}.
\newblock


\bibitem[Chalapathy and Chawla(2019)]%
        {chalapathy2019deep}
\bibfield{author}{\bibinfo{person}{Raghavendra Chalapathy} {and}
  \bibinfo{person}{Sanjay Chawla}.} \bibinfo{year}{2019}\natexlab{}.
\newblock \showarticletitle{Deep learning for anomaly detection: A survey}.
\newblock \bibinfo{journal}{\emph{arXiv preprint arXiv:1901.03407}}
  (\bibinfo{year}{2019}).
\newblock


\bibitem[Chan et~al\mbox{.}(2022)]%
        {chan2022fault}
\bibfield{author}{\bibinfo{person}{Abraham Chan}, \bibinfo{person}{Arpan
  Gujarati}, \bibinfo{person}{Karthik Pattabiraman}, {and}
  \bibinfo{person}{Sathish Gopalakrishnan}.} \bibinfo{year}{2022}\natexlab{}.
\newblock \showarticletitle{The Fault in Our Data Stars: Studying Mitigation
  Techniques against Faulty Training Data in Machine Learning Applications}. In
  \bibinfo{booktitle}{\emph{2022 52nd Annual IEEE/IFIP International Conference
  on Dependable Systems and Networks (DSN)}}. IEEE, \bibinfo{pages}{163--171}.
\newblock


\bibitem[Chen and Guestrin(2016)]%
        {chen2016xgboost}
\bibfield{author}{\bibinfo{person}{Tianqi Chen} {and} \bibinfo{person}{Carlos
  Guestrin}.} \bibinfo{year}{2016}\natexlab{}.
\newblock \showarticletitle{Xgboost: A scalable tree boosting system}. In
  \bibinfo{booktitle}{\emph{Proceedings of the 22nd acm sigkdd international
  conference on knowledge discovery and data mining}}.
  \bibinfo{pages}{785--794}.
\newblock


\bibitem[Chug et~al\mbox{.}(2021)]%
        {chug2021statistical}
\bibfield{author}{\bibinfo{person}{Sezal Chug}, \bibinfo{person}{Priya
  Kaushal}, \bibinfo{person}{Ponnurangam Kumaraguru}, {and}
  \bibinfo{person}{Tavpritesh Sethi}.} \bibinfo{year}{2021}\natexlab{}.
\newblock \showarticletitle{Statistical Learning to Operationalize a Domain
  Agnostic Data Quality Scoring}.
\newblock \bibinfo{journal}{\emph{arXiv preprint arXiv:2108.08905}}
  (\bibinfo{year}{2021}).
\newblock


\bibitem[Cichy and Rass(2019)]%
        {cichy2019overview}
\bibfield{author}{\bibinfo{person}{Corinna Cichy} {and} \bibinfo{person}{Stefan
  Rass}.} \bibinfo{year}{2019}\natexlab{}.
\newblock \showarticletitle{An overview of data quality frameworks}.
\newblock \bibinfo{journal}{\emph{IEEE Access}}  \bibinfo{volume}{7}
  (\bibinfo{year}{2019}), \bibinfo{pages}{24634--24648}.
\newblock


\bibitem[D'Agostino(2017)]%
        {d2017goodness}
\bibfield{author}{\bibinfo{person}{RalphB D'Agostino}.}
  \bibinfo{year}{2017}\natexlab{}.
\newblock \bibinfo{booktitle}{\emph{Goodness-of-fit-techniques}}.
\newblock \bibinfo{publisher}{Routledge}.
\newblock


\bibitem[Decancq and Lugo(2012)]%
        {decancq2012inequality}
\bibfield{author}{\bibinfo{person}{Koen Decancq} {and}
  \bibinfo{person}{Maria~Ana Lugo}.} \bibinfo{year}{2012}\natexlab{}.
\newblock \showarticletitle{Inequality of wellbeing: A multidimensional
  approach}.
\newblock \bibinfo{journal}{\emph{Economica}} \bibinfo{volume}{79},
  \bibinfo{number}{316} (\bibinfo{year}{2012}), \bibinfo{pages}{721--746}.
\newblock


\bibitem[Dedeke(2000)]%
        {dedeke2000conceptual}
\bibfield{author}{\bibinfo{person}{Adenekan Dedeke}.}
  \bibinfo{year}{2000}\natexlab{}.
\newblock \showarticletitle{A Conceptual Framework for Developing Quality
  Measures for Information Systems.}. In \bibinfo{booktitle}{\emph{IQ}}.
  \bibinfo{pages}{126--128}.
\newblock


\bibitem[Ehrlinger and W{\"o}{\ss}(2018)]%
        {ehrlinger2018novel}
\bibfield{author}{\bibinfo{person}{Lisa Ehrlinger} {and}
  \bibinfo{person}{Wolfram W{\"o}{\ss}}.} \bibinfo{year}{2018}\natexlab{}.
\newblock \showarticletitle{A novel data quality metric for minimality}. In
  \bibinfo{booktitle}{\emph{International Workshop on Data Quality and Trust in
  Big Data}}. Springer, \bibinfo{pages}{1--15}.
\newblock


\bibitem[Evans et~al\mbox{.}(2008)]%
        {evans2008distribution}
\bibfield{author}{\bibinfo{person}{Diane~L Evans}, \bibinfo{person}{John~H
  Drew}, {and} \bibinfo{person}{Lawrence~M Leemis}.}
  \bibinfo{year}{2008}\natexlab{}.
\newblock \showarticletitle{The distribution of the Kolmogorov--Smirnov,
  Cramer--von Mises, and Anderson--Darling test statistics for exponential
  populations with estimated parameters}.
\newblock \bibinfo{journal}{\emph{Communications in Statistics—Simulation and
  Computation{\textregistered}}} \bibinfo{volume}{37}, \bibinfo{number}{7}
  (\bibinfo{year}{2008}), \bibinfo{pages}{1396--1421}.
\newblock


\bibitem[Fan(2015)]%
        {fan2015data}
\bibfield{author}{\bibinfo{person}{Wenfei Fan}.}
  \bibinfo{year}{2015}\natexlab{}.
\newblock \showarticletitle{Data quality: From theory to practice}.
\newblock \bibinfo{journal}{\emph{Acm Sigmod Record}} \bibinfo{volume}{44},
  \bibinfo{number}{3} (\bibinfo{year}{2015}), \bibinfo{pages}{7--18}.
\newblock


\bibitem[Foidl and Felderer(2019)]%
        {foidl2019risk}
\bibfield{author}{\bibinfo{person}{Harald Foidl} {and} \bibinfo{person}{Michael
  Felderer}.} \bibinfo{year}{2019}\natexlab{}.
\newblock \showarticletitle{Risk-based data validation in machine
  learning-based software systems}. In \bibinfo{booktitle}{\emph{proceedings of
  the 3rd ACM SIGSOFT international workshop on machine learning techniques for
  software quality evaluation}}. \bibinfo{pages}{13--18}.
\newblock


\bibitem[Heinrich et~al\mbox{.}(2018a)]%
        {heinrich2018requirements}
\bibfield{author}{\bibinfo{person}{Bernd Heinrich}, \bibinfo{person}{Diana
  Hristova}, \bibinfo{person}{Mathias Klier}, \bibinfo{person}{Alexander
  Schiller}, {and} \bibinfo{person}{Michael Szubartowicz}.}
  \bibinfo{year}{2018}\natexlab{a}.
\newblock \showarticletitle{Requirements for data quality metrics}.
\newblock \bibinfo{journal}{\emph{Journal of Data and Information Quality
  (JDIQ)}} \bibinfo{volume}{9}, \bibinfo{number}{2} (\bibinfo{year}{2018}),
  \bibinfo{pages}{1--32}.
\newblock


\bibitem[Heinrich et~al\mbox{.}(2018b)]%
        {heinrich2018assessing}
\bibfield{author}{\bibinfo{person}{Bernd Heinrich}, \bibinfo{person}{Mathias
  Klier}, \bibinfo{person}{Alexander Schiller}, {and} \bibinfo{person}{Gerit
  Wagner}.} \bibinfo{year}{2018}\natexlab{b}.
\newblock \showarticletitle{Assessing data quality--A probability-based metric
  for semantic consistency}.
\newblock \bibinfo{journal}{\emph{Decision Support Systems}}
  \bibinfo{volume}{110} (\bibinfo{year}{2018}), \bibinfo{pages}{95--106}.
\newblock


\bibitem[Jain et~al\mbox{.}(2020)]%
        {jain2020overview}
\bibfield{author}{\bibinfo{person}{Abhinav Jain}, \bibinfo{person}{Hima Patel},
  \bibinfo{person}{Lokesh Nagalapatti}, \bibinfo{person}{Nitin Gupta},
  \bibinfo{person}{Sameep Mehta}, \bibinfo{person}{Shanmukha Guttula},
  \bibinfo{person}{Shashank Mujumdar}, \bibinfo{person}{Shazia Afzal},
  \bibinfo{person}{Ruhi Sharma~Mittal}, {and} \bibinfo{person}{Vitobha
  Munigala}.} \bibinfo{year}{2020}\natexlab{}.
\newblock \showarticletitle{Overview and importance of data quality for machine
  learning tasks}. In \bibinfo{booktitle}{\emph{Proceedings of the 26th ACM
  SIGKDD International Conference on Knowledge Discovery \& Data Mining}}.
  \bibinfo{pages}{3561--3562}.
\newblock


\bibitem[Jia and Harman(2010)]%
        {jia2010analysis}
\bibfield{author}{\bibinfo{person}{Yue Jia} {and} \bibinfo{person}{Mark
  Harman}.} \bibinfo{year}{2010}\natexlab{}.
\newblock \showarticletitle{An analysis and survey of the development of
  mutation testing}.
\newblock \bibinfo{journal}{\emph{IEEE transactions on software engineering}}
  \bibinfo{volume}{37}, \bibinfo{number}{5} (\bibinfo{year}{2010}),
  \bibinfo{pages}{649--678}.
\newblock


\bibitem[John et~al\mbox{.}(2021)]%
        {john2021towards}
\bibfield{author}{\bibinfo{person}{Meenu~Mary John},
  \bibinfo{person}{Helena~Holmstr{\"o}m Olsson}, {and} \bibinfo{person}{Jan
  Bosch}.} \bibinfo{year}{2021}\natexlab{}.
\newblock \showarticletitle{Towards mlops: A framework and maturity model}. In
  \bibinfo{booktitle}{\emph{2021 47th Euromicro Conference on Software
  Engineering and Advanced Applications (SEAA)}}. IEEE, \bibinfo{pages}{1--8}.
\newblock


\bibitem[Johnson et~al\mbox{.}(2016)]%
        {johnson2016application}
\bibfield{author}{\bibinfo{person}{Steven~G Johnson}, \bibinfo{person}{Stuart
  Speedie}, \bibinfo{person}{Gyorgy Simon}, \bibinfo{person}{Vipin Kumar},
  {and} \bibinfo{person}{Bonnie~L Westra}.} \bibinfo{year}{2016}\natexlab{}.
\newblock \showarticletitle{Application of an ontology for characterizing data
  quality for a secondary use of EHR data}.
\newblock \bibinfo{journal}{\emph{Applied clinical informatics}}
  \bibinfo{volume}{7}, \bibinfo{number}{01} (\bibinfo{year}{2016}),
  \bibinfo{pages}{69--88}.
\newblock


\bibitem[Kiangala and Wang(2021)]%
        {kiangala2021effective}
\bibfield{author}{\bibinfo{person}{Sonia~Kahiomba Kiangala} {and}
  \bibinfo{person}{Zenghui Wang}.} \bibinfo{year}{2021}\natexlab{}.
\newblock \showarticletitle{An effective adaptive customization framework for
  small manufacturing plants using extreme gradient boosting-XGBoost and random
  forest ensemble learning algorithms in an Industry 4.0 environment}.
\newblock \bibinfo{journal}{\emph{Machine Learning with Applications}}
  \bibinfo{volume}{4} (\bibinfo{year}{2021}), \bibinfo{pages}{100024}.
\newblock


\bibitem[Knight(2011)]%
        {knight2011combined}
\bibfield{author}{\bibinfo{person}{Shirlee-ann Knight}.}
  \bibinfo{year}{2011}\natexlab{}.
\newblock \showarticletitle{The combined conceptual life-cycle model of
  information quality: part 1, an investigative framework}.
\newblock \bibinfo{journal}{\emph{International journal of information
  quality}} \bibinfo{volume}{2}, \bibinfo{number}{3} (\bibinfo{year}{2011}),
  \bibinfo{pages}{205--230}.
\newblock


\bibitem[Kvam et~al\mbox{.}(2022)]%
        {kvam2022nonparametric}
\bibfield{author}{\bibinfo{person}{Paul Kvam}, \bibinfo{person}{Brani
  Vidakovic}, {and} \bibinfo{person}{Seong-joon Kim}.}
  \bibinfo{year}{2022}\natexlab{}.
\newblock \bibinfo{booktitle}{\emph{Nonparametric Statistics with Applications
  to Science and Engineering with R}}.
\newblock \bibinfo{publisher}{John Wiley \& Sons}.
\newblock


\bibitem[LeDell and Poirier(2020)]%
        {ledell2020h2o}
\bibfield{author}{\bibinfo{person}{Erin LeDell} {and}
  \bibinfo{person}{Sebastien Poirier}.} \bibinfo{year}{2020}\natexlab{}.
\newblock \showarticletitle{H2o automl: Scalable automatic machine learning}.
  In \bibinfo{booktitle}{\emph{Proceedings of the AutoML Workshop at ICML}},
  Vol.~\bibinfo{volume}{2020}.
\newblock


\bibitem[Lee(2000)]%
        {lee2000measures}
\bibfield{author}{\bibinfo{person}{Lillian Lee}.}
  \bibinfo{year}{2000}\natexlab{}.
\newblock \showarticletitle{Measures of distributional similarity}.
\newblock \bibinfo{journal}{\emph{arXiv preprint cs/0001012}}
  (\bibinfo{year}{2000}).
\newblock


\bibitem[Li and Offutt(2016)]%
        {li2016test}
\bibfield{author}{\bibinfo{person}{Nan Li} {and} \bibinfo{person}{Jeff
  Offutt}.} \bibinfo{year}{2016}\natexlab{}.
\newblock \showarticletitle{Test oracle strategies for model-based testing}.
\newblock \bibinfo{journal}{\emph{IEEE Transactions on Software Engineering}}
  \bibinfo{volume}{43}, \bibinfo{number}{4} (\bibinfo{year}{2016}),
  \bibinfo{pages}{372--395}.
\newblock


\bibitem[Lin(1991)]%
        {lin1991divergence}
\bibfield{author}{\bibinfo{person}{Jianhua Lin}.}
  \bibinfo{year}{1991}\natexlab{}.
\newblock \showarticletitle{Divergence measures based on the Shannon entropy}.
\newblock \bibinfo{journal}{\emph{IEEE Transactions on Information theory}}
  \bibinfo{volume}{37}, \bibinfo{number}{1} (\bibinfo{year}{1991}),
  \bibinfo{pages}{145--151}.
\newblock


\bibitem[Lionis et~al\mbox{.}(2021)]%
        {lionis2021rssi}
\bibfield{author}{\bibinfo{person}{Antonios Lionis},
  \bibinfo{person}{Konstantinos~P Peppas}, \bibinfo{person}{Hector~E
  Nistazakis}, {and} \bibinfo{person}{Andreas Tsigopoulos}.}
  \bibinfo{year}{2021}\natexlab{}.
\newblock \showarticletitle{RSSI Probability Density Functions Comparison Using
  Jensen-Shannon Divergence and Pearson Distribution}.
\newblock \bibinfo{journal}{\emph{Technologies}} \bibinfo{volume}{9},
  \bibinfo{number}{2} (\bibinfo{year}{2021}), \bibinfo{pages}{26}.
\newblock


\bibitem[Loshin(2010)]%
        {loshin2010practitioner}
\bibfield{author}{\bibinfo{person}{David Loshin}.}
  \bibinfo{year}{2010}\natexlab{}.
\newblock \bibinfo{booktitle}{\emph{The practitioner's guide to data quality
  improvement}}.
\newblock \bibinfo{publisher}{Elsevier}.
\newblock


\bibitem[Lu et~al\mbox{.}(2018)]%
        {lu2018learning}
\bibfield{author}{\bibinfo{person}{Jie Lu}, \bibinfo{person}{Anjin Liu},
  \bibinfo{person}{Fan Dong}, \bibinfo{person}{Feng Gu}, \bibinfo{person}{Joao
  Gama}, {and} \bibinfo{person}{Guangquan Zhang}.}
  \bibinfo{year}{2018}\natexlab{}.
\newblock \showarticletitle{Learning under concept drift: A review}.
\newblock \bibinfo{journal}{\emph{IEEE Transactions on Knowledge and Data
  Engineering}} \bibinfo{volume}{31}, \bibinfo{number}{12}
  (\bibinfo{year}{2018}), \bibinfo{pages}{2346--2363}.
\newblock


\bibitem[Martens and Martens(2001)]%
        {martens2001multivariate}
\bibfield{author}{\bibinfo{person}{Harald Martens} {and} \bibinfo{person}{Magni
  Martens}.} \bibinfo{year}{2001}\natexlab{}.
\newblock \bibinfo{booktitle}{\emph{Multivariate analysis of quality: an
  introduction}}.
\newblock \bibinfo{publisher}{John Wiley \& Sons}.
\newblock


\bibitem[Meng and Ci(2013)]%
        {meng2013big}
\bibfield{author}{\bibinfo{person}{Xiaofeng Meng} {and} \bibinfo{person}{Xiang
  Ci}.} \bibinfo{year}{2013}\natexlab{}.
\newblock \showarticletitle{Big data management: concepts, techniques and
  challenges}.
\newblock \bibinfo{journal}{\emph{Journal of computer research and
  development}} \bibinfo{volume}{50}, \bibinfo{number}{1}
  (\bibinfo{year}{2013}), \bibinfo{pages}{146--169}.
\newblock


\bibitem[Moges et~al\mbox{.}(2013)]%
        {moges2013multidimensional}
\bibfield{author}{\bibinfo{person}{Helen-Tadesse Moges}, \bibinfo{person}{Karel
  Dejaeger}, \bibinfo{person}{Wilfried Lemahieu}, {and} \bibinfo{person}{Bart
  Baesens}.} \bibinfo{year}{2013}\natexlab{}.
\newblock \showarticletitle{A multidimensional analysis of data quality for
  credit risk management: New insights and challenges}.
\newblock \bibinfo{journal}{\emph{Information \& Management}}
  \bibinfo{volume}{50}, \bibinfo{number}{1} (\bibinfo{year}{2013}),
  \bibinfo{pages}{43--58}.
\newblock


\bibitem[Muller et~al\mbox{.}(2019)]%
        {muller2019data}
\bibfield{author}{\bibinfo{person}{Michael Muller}, \bibinfo{person}{Ingrid
  Lange}, \bibinfo{person}{Dakuo Wang}, \bibinfo{person}{David Piorkowski},
  \bibinfo{person}{Jason Tsay}, \bibinfo{person}{Q~Vera Liao},
  \bibinfo{person}{Casey Dugan}, {and} \bibinfo{person}{Thomas Erickson}.}
  \bibinfo{year}{2019}\natexlab{}.
\newblock \showarticletitle{How data science workers work with data: Discovery,
  capture, curation, design, creation}. In
  \bibinfo{booktitle}{\emph{Proceedings of the 2019 CHI conference on human
  factors in computing systems}}. \bibinfo{pages}{1--15}.
\newblock


\bibitem[Nguyen and Vreeken(2015)]%
        {nguyen2015non}
\bibfield{author}{\bibinfo{person}{Hoang-Vu Nguyen} {and}
  \bibinfo{person}{Jilles Vreeken}.} \bibinfo{year}{2015}\natexlab{}.
\newblock \showarticletitle{Non-parametric jensen-shannon divergence}. In
  \bibinfo{booktitle}{\emph{Joint European conference on machine learning and
  knowledge discovery in databases}}. Springer, \bibinfo{pages}{173--189}.
\newblock


\bibitem[Olson(2003)]%
        {olson2003data}
\bibfield{author}{\bibinfo{person}{Jack~E Olson}.}
  \bibinfo{year}{2003}\natexlab{}.
\newblock \bibinfo{booktitle}{\emph{Data quality: the accuracy dimension}}.
\newblock \bibinfo{publisher}{Elsevier}.
\newblock


\bibitem[Peng and Lei(2005)]%
        {peng2005review}
\bibfield{author}{\bibinfo{person}{Liu Peng} {and} \bibinfo{person}{Lei Lei}.}
  \bibinfo{year}{2005}\natexlab{}.
\newblock \showarticletitle{A review of missing data treatment methods}.
\newblock \bibinfo{journal}{\emph{Intell. Inf. Manag. Syst. Technol}}
  \bibinfo{volume}{1} (\bibinfo{year}{2005}), \bibinfo{pages}{412--419}.
\newblock


\bibitem[Pipino et~al\mbox{.}(2002)]%
        {pipino2002data}
\bibfield{author}{\bibinfo{person}{Leo~L Pipino}, \bibinfo{person}{Yang~W Lee},
  {and} \bibinfo{person}{Richard~Y Wang}.} \bibinfo{year}{2002}\natexlab{}.
\newblock \showarticletitle{Data quality assessment}.
\newblock \bibinfo{journal}{\emph{Commun. ACM}} \bibinfo{volume}{45},
  \bibinfo{number}{4} (\bibinfo{year}{2002}), \bibinfo{pages}{211--218}.
\newblock


\bibitem[Pratt and Gibbons(2012)]%
        {pratt2012concepts}
\bibfield{author}{\bibinfo{person}{John~Winsor Pratt} {and}
  \bibinfo{person}{Jean~Dickinson Gibbons}.} \bibinfo{year}{2012}\natexlab{}.
\newblock \bibinfo{booktitle}{\emph{Concepts of nonparametric theory}}.
\newblock \bibinfo{publisher}{Springer Science \& Business Media}.
\newblock


\bibitem[Rettig et~al\mbox{.}(2019)]%
        {rettig2019online}
\bibfield{author}{\bibinfo{person}{Laura Rettig}, \bibinfo{person}{Mourad
  Khayati}, \bibinfo{person}{Philippe Cudr{\'e}-Mauroux}, {and}
  \bibinfo{person}{Micha{\l} Pi{\'o}rkowski}.} \bibinfo{year}{2019}\natexlab{}.
\newblock \showarticletitle{Online anomaly detection over big data streams}.
\newblock In \bibinfo{booktitle}{\emph{Applied data science}}.
  \bibinfo{publisher}{Springer}, \bibinfo{pages}{289--312}.
\newblock


\bibitem[Schuler and Zeller(2013)]%
        {schuler2013covering}
\bibfield{author}{\bibinfo{person}{David Schuler} {and}
  \bibinfo{person}{Andreas Zeller}.} \bibinfo{year}{2013}\natexlab{}.
\newblock \showarticletitle{Covering and uncovering equivalent mutants}.
\newblock \bibinfo{journal}{\emph{Software Testing, Verification and
  Reliability}} \bibinfo{volume}{23}, \bibinfo{number}{5}
  (\bibinfo{year}{2013}), \bibinfo{pages}{353--374}.
\newblock


\bibitem[Sunderland et~al\mbox{.}(2019)]%
        {sunderland2019utility}
\bibfield{author}{\bibinfo{person}{Kelly~M Sunderland}, \bibinfo{person}{Derek
  Beaton}, \bibinfo{person}{Julia Fraser}, \bibinfo{person}{Donna Kwan},
  \bibinfo{person}{Paula~M McLaughlin}, \bibinfo{person}{Manuel
  Montero-Odasso}, \bibinfo{person}{Alicia~J Peltsch},
  \bibinfo{person}{Frederico Pieruccini-Faria}, \bibinfo{person}{Demetrios~J
  Sahlas}, \bibinfo{person}{Richard~H Swartz}, {et~al\mbox{.}}}
  \bibinfo{year}{2019}\natexlab{}.
\newblock \showarticletitle{The utility of multivariate outlier detection
  techniques for data quality evaluation in large studies: an application
  within the ONDRI project}.
\newblock \bibinfo{journal}{\emph{BMC medical research methodology}}
  \bibinfo{volume}{19}, \bibinfo{number}{1} (\bibinfo{year}{2019}),
  \bibinfo{pages}{1--16}.
\newblock


\bibitem[Taleb et~al\mbox{.}(2021)]%
        {taleb2021big}
\bibfield{author}{\bibinfo{person}{Ikbal Taleb}, \bibinfo{person}{Mohamed~Adel
  Serhani}, \bibinfo{person}{Chafik Bouhaddioui}, {and}
  \bibinfo{person}{Rachida Dssouli}.} \bibinfo{year}{2021}\natexlab{}.
\newblock \showarticletitle{Big data quality framework: a holistic approach to
  continuous quality management}.
\newblock \bibinfo{journal}{\emph{Journal of Big Data}} \bibinfo{volume}{8},
  \bibinfo{number}{1} (\bibinfo{year}{2021}), \bibinfo{pages}{1--41}.
\newblock


\bibitem[Teh et~al\mbox{.}(2020)]%
        {teh2020sensor}
\bibfield{author}{\bibinfo{person}{Hui~Yie Teh}, \bibinfo{person}{Andreas~W
  Kempa-Liehr}, {and} \bibinfo{person}{Kevin I-Kai Wang}.}
  \bibinfo{year}{2020}\natexlab{}.
\newblock \showarticletitle{Sensor data quality: A systematic review}.
\newblock \bibinfo{journal}{\emph{Journal of Big Data}} \bibinfo{volume}{7},
  \bibinfo{number}{1} (\bibinfo{year}{2020}), \bibinfo{pages}{1--49}.
\newblock


\bibitem[Vaziri et~al\mbox{.}(2019)]%
        {vaziri2019measuring}
\bibfield{author}{\bibinfo{person}{Reza Vaziri}, \bibinfo{person}{Mehran
  Mohsenzadeh}, {and} \bibinfo{person}{Jafar Habibi}.}
  \bibinfo{year}{2019}\natexlab{}.
\newblock \showarticletitle{Measuring data quality with weighted metrics}.
\newblock \bibinfo{journal}{\emph{Total Quality Management \& Business
  Excellence}} \bibinfo{volume}{30}, \bibinfo{number}{5-6}
  (\bibinfo{year}{2019}), \bibinfo{pages}{708--720}.
\newblock


\bibitem[Wand and Wang(1996)]%
        {wand1996anchoring}
\bibfield{author}{\bibinfo{person}{Yair Wand} {and} \bibinfo{person}{Richard~Y
  Wang}.} \bibinfo{year}{1996}\natexlab{}.
\newblock \showarticletitle{Anchoring data quality dimensions in ontological
  foundations}.
\newblock \bibinfo{journal}{\emph{Commun. ACM}} \bibinfo{volume}{39},
  \bibinfo{number}{11} (\bibinfo{year}{1996}), \bibinfo{pages}{86--95}.
\newblock


\bibitem[Wang et~al\mbox{.}(1995)]%
        {wang1995framework}
\bibfield{author}{\bibinfo{person}{Richard~Y Wang}, \bibinfo{person}{Veda~C
  Storey}, {and} \bibinfo{person}{Christopher~P Firth}.}
  \bibinfo{year}{1995}\natexlab{}.
\newblock \showarticletitle{A framework for analysis of data quality research}.
\newblock \bibinfo{journal}{\emph{IEEE transactions on knowledge and data
  engineering}} \bibinfo{volume}{7}, \bibinfo{number}{4}
  (\bibinfo{year}{1995}), \bibinfo{pages}{623--640}.
\newblock


\end{thebibliography}


\end{document}